# Selection of model in developing Information Security criteria on Smart Grid Security System Paper

Amy Poh Ai Ling
GCOE Program, MIMS
Graduate School of Science and Technology,
Meiji University, Kanagawa-Ken, Japan
amypoh@meiji.ac.jp

Mukaidono Masao
Computer Science Department
School of Science and Technology, Meiji University,
Kanagawa-Ken, Japan
masao@cs.meiji.ac.jp

*Abstract*— **At present, "Smart Grid" emerged to be one of the best advanced energy supply chain. This paper looks into the security system of Smart Grid via Smart Planet system. The scope focused onto information security criteria that impacts consumer trust and satisfaction. The importance of information security criteria is the main aspect perceived to impact customer trust towards the entire smart grid system. On one hand, it also focused on the selection of the model in developing information security criteria on smart grid.**

## I. INTRODUCTION

As the consumer's awareness and participation on smart grid project has increased, criteria that hold the ability to protect the information security will be the main issue. This will impacts the satisfaction that consumer perceive and trust towards the entire system. This paper identified the criteria that could enhance information security system of a smart grid project.

Smart grid features intelligent monitoring of the status and amounts of the electricity flowing throughout the grid, [1] it is perhaps the "cloud computing" of the utility industry. However, the increasing use of IT-based electric power systems increases cyber security vulnerabilities, and this increases the importance of cyber security. [2] Thus, the IT and information security industries need to pay more attention to the electricity grid in the future as more and more smart grids are set up with two-way communication systems. [3]

It is important to select a good model as a starting point. This paper breaks into three sections: First, the selection of three models, one of which is 'privileged', taking into consideration Quality Function Deployment (QFD) versus Object Oriented (OO), Joint Application Design (JAD), Cleanroom, Structured Analysis and Structured Design (SASD) and ConJoint Analysis (CA); Second, the methods and techniques to support the selected models; Third, the information security criteria identification.

## II. METHODOLOGY

The focus of this project was to identify the importance of information security in the targeted area. This was done through two steps: Firstly, through expert group discussion and study models comparison, concentrate on the selection of the appropriate model to be applied for the project; secondly, study on the methods and techniques to support the selected model. By literature review and experts opinion focused group discussion the information security criteria was identified.

### A. Project Focused

In January, 2010, IBM Chairman Sam Palmisano addressed Chatham House in London, where he described how forward-thinking leaders in business, government and civil society around the world are capturing the potential of smarter systems to achieve economic growth, near-term efficiency, sustainable development and societal progress. [4] Looking very interestingly into Smart Grid via Smart Planet system, it was found that information security's impact on customer trust and satisfaction is a valuable topic.

### B. Conceptual Mind Mapping

Over the past several years, the promise of smart grids and their benefits has been widely publicized. Bringing updated technologies to power generation, transmission, and consumption, smart grids are touted to revolutionize economy, environment, and national security.

Figure 1 below shows Smart grid conceptual mapping where the study area is highlighted. Under the umbrella of Smart Planet cultivated by IBM, lay Smart Water Management and Green Planet, Smarter Planet Skill and Education, Smart Grid, Smart Health Care and Smart Cities. There are eight main elements in Smart Grid, narrowing down to network security utilities, business network, communication and information security's impact on consumer trust and satisfaction which is the main focus of this paper. Smart grid utilizes communication technology and information to optimally transmit and distribute electricity from suppliers to consumers. Alternatively, security professionals immediately consider the new risks that these new functionalities and benefits have introduced to the environment [16].

### C. Importance of Information Security

The fundamental components to a Smart Grid are real time monitoring and reaction of events on the Grid, anticipation of upcoming events, and problem isolation. In order to realize the Smart Grid, data are gathered from large numbers of intelligent sensors and processors installed on the power lines and equipment of the distribution grid. These data collected will be transferred to central information processing systems which both present the information to operators and use the information to send back control settings. [17]

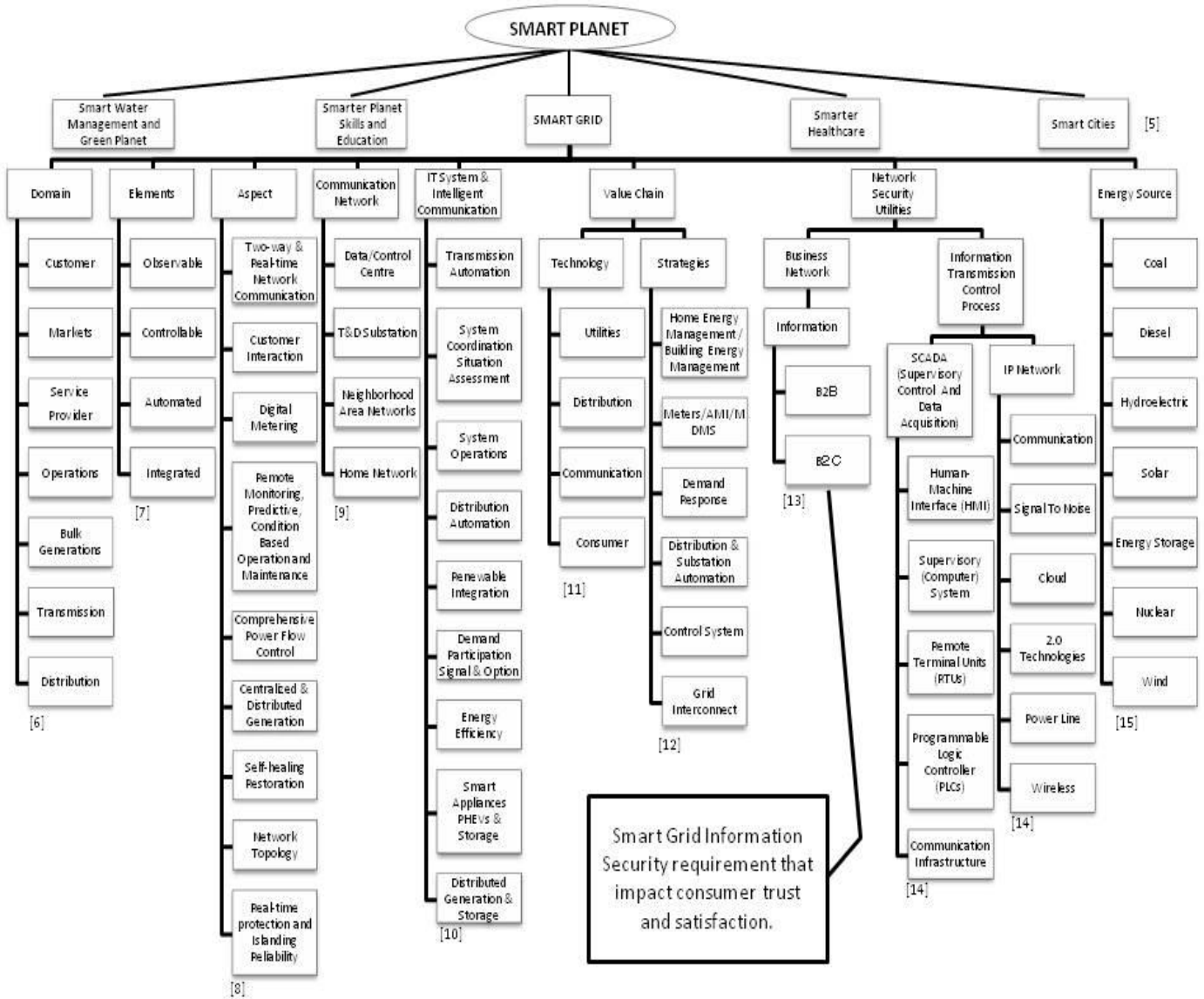

Figure 1. Smart Planet Conceptual Mapping

While information and communication seems to be much more importance compare to decades ago, the system has to be mightily strong to protect itself against hacker's attack.

When the information is share real-time between power generator, distributed resources, service provider, control center, substation, even to end-users, any changes expose to hacker's attack would bring the whole system down to mess. This will dangerously create consumer distrust and dissatisfaction that may lead to other more destructive phenomenon.

The importance of information security and the information security's impact on consumer trust and satisfaction mind mapping was portrayed in Figure 2.

Critical infrastructure systems are dramatically increasing in complexity. [19] Contrast that with the minute level of detail we are used to get from network systems, security, and performance tuning tools in the IT world, we begin to appreciate the degree of blindness we accept with regards to accurately measuring and managing our electricity consumption. [20]

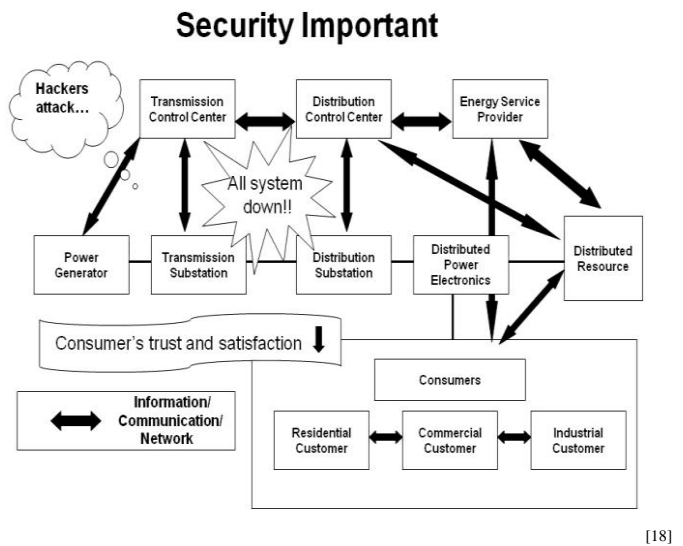

Figure 2. Information Security's Impact on Consumer Trust and Satisfaction

## D. Problem Statement

Smart planet conceptual mapping demonstrate the linkage between business to consumer (B2C). The success of this effort appeared to hinge on utility companies championing information system security initiatives and propagating an awareness of the importance of information security among consumer at all levels of the community. Hence, an interesting inquiry developed looking at the Smart Grid Information Security requirements that impact consumer trust and satisfaction. What are the requirements that would bring such impact, and what is the significance each requirement bears? This paper will discuss about the impact and significance of each requirements identified.

## III. MODEL SELECTION

### A. Data Analysis And Development Models Comparison

Five major data analysis and development models were picked for models study and comparison to select appropriate methodology model to identify information security criteria for this paper. Table I shows QFD versus Object Oriented (OO), Joint Application Design (JAD), Cleanroom, Structured Analysis and Structured Design (SASD) and ConJoint Analysis (CA) looking from different perspective.

TABLE I. QFD VERSUS OO, JAD, CLASSROOM, SASD, CA

| OO | JAD | Clean room | SASD | CA |
|---|---|---|---|---|
| QFD focuses on the quality of the system. OO focuses on good software design. | Customer requirements are the driving force of QFD. Driving force for JAD is human communication. | Cleanroom focuses on producing error-free system. QFD focuses on customer satisfaction and competitive performance. | QFD uses HOQ to produce a prioritization of requirements. SASD doesn't use prioritization. | CA is more efficient in reflecting the end-users' present preferences for the product attributes. QFD is definitely better in satisfying end-users' needs from the developers point of view. |
| QFD requirements have to be firm and clear before starting. OO can work with changing requirements | QFD main element: HOQ. JAD main element: proper communication. | QFD helps elicit and build the customer requirements. Cleanroom starts after the elicitation of requirements. | QFD requires user participation in almost all the stages. SASD requires user participation in requirements elicitation only. | CA is easier to compare the most preferred features to profit maximizing features and also to develop designs that optimize product line sales or profits. |
| OO changing requirements do cost more and requires more software life cycle iterations. | QFD focuses on quality. JAD focuses on structural analysis. | Cleanroom utilizes formal specifications. QFD uses natural languages. | SASD takes us a step further than QFD into building an implementation model. | QFD is able to highlight the fact that certain engineering characteristics or design features had both positive and negative aspects. |
| QFD bypasses this cost by having the requirements clear at the initial stages. | QFD takes longer for results, requires commitment from both analysts and users in all phases. JAD requires appropriate human communication between the analysts and users. | QFD focuses on the voice of the customer and customer satisfaction while Cleanroom is more focused on error prevention. | QFD is used mostly for competitor analysis for generic products, while SASD is used to build user-specific systems. | QFD also highlighted the importance of starting explicitly with customer needs, regardless of which method is used. QFD's ability to generate creative or novel solutions could be combined with conjoint analysis' ability to forecast market reaction to design changes. |
| Area focus: Software | Area focus: Human Communication | Area focus: Error Prevention | Area focus: Build User-Specific Systems | Area focus: Market Forecast |

[21], [22], [23], [24]

QFD consider being more appropriate model in developing information security criteria as it gain its credit focuses on the quality of the system, bypassing changing requirements and software compared to OO model.

In terms comparing with JAD, QFD has its privilege in this project for having customer requirements as its driving force. On the other hand, QFD versus Cleanroom explicit QFD's strong points in focuses more on customer satisfaction and competitive performance, QFD also helps elicit and build customer requirements by using natural language.

Contradict with SASD, QFD uses House of Quality (HOQ) to produce a prioritization of requirements where as SASD doesn't use prioritization which seems to be an important step in developing criteria. Nonetheless, QFD take a step forward compare to CA with its good ability of satisfying end-user's need from the developer point of view, QFD also highlighted the importance of starting explicitly with customer needs, regardless of which method is used.

Therefore, in conclusion, we employ QFD as our methodology model.

### B. Quality Function Deployment (QFD)

QFD is a method to transform user demands into design quality, to deploy the functions forming quality, and to deploy methods for achieving the design quality into subsystems and component parts, and ultimately to specific elements of the manufacturing process. QFD House of Quality (HOQ) is a matrix consists of a series of rows and columns. Each row reflects a clear business objective, and each column reflects a separate mechanism, in this paper it portray information security requirements that a consumer perceived. See Figure 3 below.

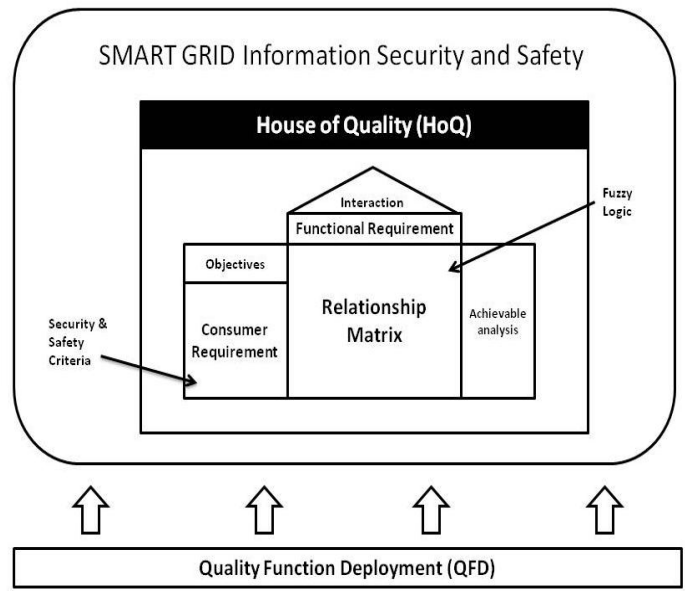

Figure 3. Quality Function Deployment

The cells represent the capacity for each requirement mechanism to influence the attainment of each objective. Developing the framework for the matrix provides an opportunity for the authority involved to reconsider the system fundamental information security mission, its logic, and its operating structure.

Through a simple mathematical relationship, QFD provides a means of ranking the objectives, and prioritizing improvement of the requirements identified. By further developing the basic diagram, a triangular, half-matrix on top of the main QFD diagram provides an opportunity to look at the interactions between each of the functional requirements and to highlight the extent to which they are likely to work in harmony or conflict.

*C. Function of QFD*

QFD is a method to transform users' desire into design quality, to deploy the functions forming quality, and to deploy methods for achieving the design quality into subsystems, and ultimately to specific elements of a specified process. In this paper, QFD identify critical customer attributes such as consumer requirements and creating a specific link between customers attributes design parameters.

QFD matrices are used to organize information safety requirements to help involved authority and design engineers identify attributes that are critical to customers and important design parameters in driving those customer attributes.

*D. Benefit of using QFD tool*

QFD allows the creation of multifunctional model that include multiple points of consideration and significance of the study. In this paper, QFD typically aims to complete the determination of specific improvement work aims for Smart Grid Information Security and Safety and displaying of a wide variety of important design information in one place in a compact form.

IV. METHODS AND TECHNIQUES TO SUPPORT THE SELECTED MODELS

Studies indicate that as much as somewhere between 35 per cent and 44 per cent of all products launched is considered failures [25]. QFD provides visual connective process to focus on the needs of the customers, helps to develop more customer-oriented, higher-quality products.

While the structure provided by QFD can be significantly beneficial, it needs techniques such as fuzzy logic, artificial neural networks, and the Taguchi method to be combined in resolving some of its drawbacks.

The merging of these techniques is envisaged to make the QFD process more robust, more quantitatively-oriented, and bring together the different stages of the QFD process [26].

The application of fuzzy logic theory provides a more quantitative method in evaluating the subjective decision-making process in the QFD analysis. On the other hand, a machine-learning approach, using Artificial Neural Network (ANN), has been suggested to resolve part of this problem by computing the customer's satisfaction index objectively instead of the customer subjectively ranking the competitors' and in-house products in the customer evaluation part of the house of quality.

The Taguchi method has been proposed to help benchmark in the house of quality with its specialty combination of an engineering approach and a statistical method to achieve improvements in product or process's cost and quality, accomplished through design optimization. QFD identifies the direction of improvement for certain design parameters, but cannot give the exact amount of improvements or the exact target values. The Taguchi method will help determine target values for the manufacturing process. The study of these three techniques were portray in Table II below.

TABLE II. METHODS AND TECHNIQUES TO HELP QUALITY FUNCTION DEPLOYMENT (QFD)

| Fuzzy Logic | Artificial Neural Network (ANN) | Taguchi Method |
|---|---|---|
| To treat or permit the uncertainty in the data collected. | Considered as simplified mathematical models of the human brain which function as computing networks (Hammerstrom, 1993). | A combination of an engineering approach and a statistical method to achieve improvements in product/process's cost and quality, accomplished through design optimization. |
| Can model vagueness in data and/or relationship in a formal way. | Makes use of the way that the human brain learns and functions and represents this information in mathematical algorithms incorporated in computers. | Is to identify parameters that can be controlled (control factors) and to reduce the sensitivity of engineering designs to uncontrollable factors (noise). |
| Able to manipulate fuzzy qualitative data in terms of linguistic variables. | Ability to learn from examples and thus have the ability to manage systems from their observed behavior. | Achieved by using small-scale experiments in the laboratory to find reliable designs for large-scale production. |
| Fuzzy Logic uses human linguistic understanding to express the knowledge of a system: facts, concepts, theories, procedures and relationships and is expressed in the form of IF-THEN rules. | Ability to learn from experience is very useful in the real world. | Three major contributions to the field of quality (Taguchi, 1993):<br>(1) the quality loss function;<br>(2) orthogonal arrays; and<br>(3) robustness |
| Fuzzy logic exhibits some useful features for exploitation in QFD:<br>-uses human linguistic understanding to express the knowledge of the systems.<br>-allows decision making with estimated values under incomplete or uncertain information.<br>-suitable for uncertain or approximate reasoning.<br>-interpretation of its | ANN exhibits some valuable features that can be useful for merging it with QFD:<br>-ability to deal with a large amount of input data.<br>-ability to deal with imprecise data and ill-defined activities ± they can tolerate faults.<br>- adaptive, possessing the | Some of the benefits of the Taguchi method can prove useful for exploitation in QFD:<br>-the modeling of interactions between characteristics, useful for the roof area of the house of quality;<br>-the optimization of target values using the loss function, useful for setting target values after customer and technical benchmarking |
| rules is simple and easy to understand; and<br>-Deals with multi-input, multi-output systems. | ability to learn from examples.<br>-can reduce development time by learning underlying relationships;<br>-non-linear, can capture complex interactions among the input variables in a system. | in QFD;<br>-determining the nature of relationships between demands and optimize the conflicts; and<br>-helping to design robust products that are insensitive to variations in environmental conditions. |
| Fuzzy logic has the ability to deal with subjective decisions and is particularly suited as a quantitative method to evaluate these subjective decision-making processes. Fuzzy gives details on how to construct an overall customer satisfaction index to determine the best product among the competitors based on the use of the technique for order preference by similarity to ideal solution (TOPSIS). | The ability of the neural network to generalize functional relationships among example data is of great importance for design. This property is important wherever these functional relationships are assumed, but not known. | Taguchi's philosophy of robust design is for establishing the best operating conditions for manufacturing and can thus be integrated in the third QFD phase, the process planning phase. Taguchi's quality loss function offers an improved way to accomplish technical benchmarking at the bottom of the HOQ. Before benchmarking, the team is really dealing with customer perceptions and not actual performance. |

[26]

In this case, fuzzy logic seems to be more efficient to reduce the uncertainty in the data collected, model vagueness in data and relationship in a formal way and manipulate fuzzy qualitative data in terms of linguistic variables.

Fuzzy logic allows decision making with estimated values under incomplete or uncertain information and has the ability to deal with subjective decisions and is particularly suited as a quantitative method to evaluate these subjective decision-making processes.

## V. INFORMATION SECURITY REQUIREMENTS IDENTIFICATION

### A. Requirement Identification

In identifying information security requirement, the significant relationship to consumer's need is taken into consideration as how it would impact consumer's trust and satisfaction, as showed in Table III.

The identified amount of sixteen requirement after scanning via focused group discussion are Confidentiality, Integrity, Availability, Device level, Cryptography and key management, Systems level, Networking issues, Strategic support, Quality assurance, Tactical oversight, Privacy concern, Low bandwidth of communications channels, Microprocessor constraints on memory and compute capabilities, Wireless media, Immature or proprietary protocols and Facilities misuse prevention. These are the core criteria of information security that must be applied to ensure the smart grid goals are achieved, thus create consumer trust and satisfaction towards the entire Smart Grid to Smart Planet system.

TABLE III. INFORMATION SECURITY CRITERIA IDENTIFICATION

| $(Cr_{(j\_n)})$ | Consumer Requirement | Significance |
|---|---|---|
| $(Cr_{(j\_1)})$ | Confidentiality | Unauthorized disclosure of information. |
| $(Cr_{(j\_2)})$ | Integrity | Unauthorized modification or destruction of information; Strong requirement that information should not be modified by unauthorized entities, and should be validated for accuracy and errors. |
| $(Cr_{(j\_3)})$ | Availability | Disruption of access to or use of information or an information system; Strong requirement that information should be available within appropriate time frames. |
| $(Cr_{(j\_4)})$ | Device level | Efforts to devise cost-effective, tamper-resistant architectures for smart meters and other components, which are necessary for systems-level survivability and resiliency and for improving intrusion detection in embedded systems. |
| $(Cr_{(j\_5)})$ | Cryptography and key management | To enable key management on a scale involving, potentially, tens of millions of credentials and keys as well as local cryptographic process on the sensors such as encryption and digital signatures. |
| $(Cr_{(j\_6)})$ | Systems level | Where research on a number of related topics is required to further approaches to building advanced protection architecture that can evolve and can tolerate failures, perhaps of a significant subset of constituents. |
| $(Cr_{(j\_7)})$ | Networking issues | Investigate ways to ensure that commercially available components, public networks like the Internet, or available enterprise systems can be implemented without jeopardizing security or reliability. |
| $(Cr_{(j\_8)})$ | Strategic support | Assessments of security properties can be used to aid different kinds of decision making, such as program planning, resource allocation, and product and service selection. |
| $(Cr_{(j\_9)})$ | Quality assurance | Security metrics can be used during the software development lifecycle to eliminate vulnerabilities, particularly during code production, by performing functions such as measuring adherence to secure coding standards, identifying likely vulnerabilities. |
| $(Cr_{(j\_10)})$ | Tactical oversight | Monitoring and reporting of the security status or posture of an IT system can be carried out to determine compliance with security requirements (e.g., policy, procedures, and regulations), gauge the effectiveness of security controls and manage risk. |
| $(Cr_{(j\_11)})$ | Privacy concern | Strong requirement that information should not be viewed by unauthorized entities. |
| $(Cr_{(j\_12)})$ | Low bandwidth of communications channels | Severely-limited bandwidth may constrain the types of security technologies that should be used across an interface while still meeting that interface's performance requirements. |
| (Cr_(j_13)) | Microprocessor constraints on memory and compute capabilities | Severely-limited memory and/or compute capabilities of a microprocessor-based platform may constrain the types of security technologies, such as cryptography, that may be used while still allowing the platform to meet its performance requirements. |
| (Cr_(j_14)) | Wireless media | Wireless media may necessitate specific types of security technologies to address wireless vulnerabilities across the wireless path. |
| (Cr_(j_15)) | Immature or proprietary protocols | Immature or proprietary protocols may not be adequately tested either against inadvertent compromises or deliberate attacks. This may leave the interface with more vulnerabilities than if a more mature protocol were used. |
| (Cr_(j_16)) | Facilities misuse prevention | Organization monitored for unauthorized use of information processing facilities, users aware of their exact scope of permitted access. Users aware that monitoring tools are being used to detect unauthorized use. |

[27], [28], [29], [30]

With proper security controls, smart grids are able to either prevent or minimize the negative impact of hackers attack; thus, increasing the reliability of the grid, gaining trust and satisfaction of the users. Security can either be viewed as an obstacle to the progress of smart grids or as an enabler to allow nations to meet their ambitious smart grid goals. [17]

*B. Identified Consumer Requirements*

| Row # | | Consumer Requirements (Explicit and Implicit) |
|---|---|---|
| 1 | (Cr_(j_1)) | Confidentiality |
| 2 | (Cr_(j_2)) | Integrity |
| 3 | (Cr_(j_3)) | Availability |
| 4 | (Cr_(j_4)) | Device level |
| 5 | (Cr_(j_5)) | Cryptography and key management |
| 6 | (Cr_(j_6)) | Systems level |
| 7 | (Cr_(j_7)) | Networking issues |
| 8 | (Cr_(j_8)) | Strategic support |
| 9 | (Cr_(j_9)) | Quality assurance |
| 10 | (Cr_(j_10)) | Tactical oversight |
| 11 | (Cr_(j_11)) | Privacy concern |
| 12 | (Cr_(j_12)) | Low bandwidth of communications channels |
| 13 | (Cr_(j_13)) | Microprocessor constraints on memory and compute capabilities |
| 14 | (Cr_(j_14)) | Wireless media |
| 15 | (Cr_(j_15)) | Immature or proprietary protocols |
| 16 | (Cr_(j_16)) | Facilities misuse prevention |

Figure 4. QFD House of Quality (HOQ) Left Column

Identified information security requirements were inserted into left column of the House of Quality, show in Figure 4.

The sixteen identified consumer requirements serve as the key to consumer empowerment while no data transmission over the Internet or information storage technology can be guaranteed to be 100% secure, and the sensitivity of the customer information at issue.

## VI. CONCLUSION

This paper provide sufficient information for the reader to understand the basic conceptual mapping for a smart planet implementation and to summarize the smart grid system at a standard level as focused on the information security requirements, without documenting the use cases to the full level of documentation detail. Focused area was illustrated with the developed of Smart Grid conceptual mind mapping. Hence, the importance of information security in a smart grid system and how it would bring direct impact to consumer trust and satisfaction was elucidated.

The comparison to select appropriate methodology model was done with the elements of QFD versus OO, JAD, Classroom, SASD and CA looking at different perspective. In addition, three methods and techniques were taken into consideration to enhance development of QFD House of Quality, fuzzy logic won out as a quantitative method to evaluate subjective decision-making processes.

The information security requirements were carefully picked from literature review, referring experts' opinion and focused group discussion before inserting into the left column of HOQ. This pave a smooth way for further research development on the HOQ with the insertion of information security functional requirements in the top column of HOQ to fulfill the criteria, and fuzzy logic value to develop an information security relationship matrix to understand criteria's significant.

## VII. FUTURE RESEARCH DIRECTION

We are now working on the identification and allocation of information security functional requirements that intend to fulfill the criteria mentioned in Figure 5.

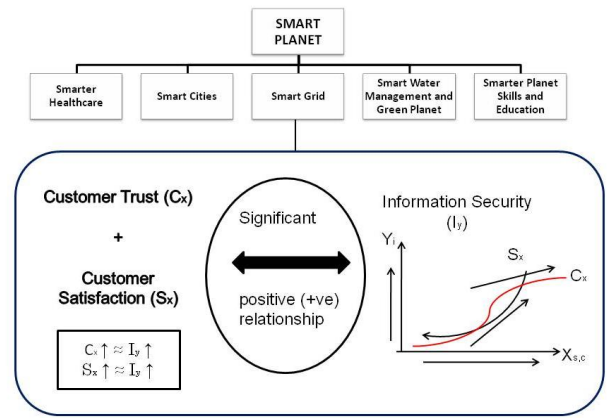

Figure 5. Relationship Proofing between Information Security with Consumern Trust and Satisfaction

It is targeted to contribute towards the further development of the HOQ of this paper.

Figure 5 elucidate the relationship between information security with customer trust and satisfaction. We design to prove that there is a significant positive relationship between the said elements with the help of fuzzy logic value which would serve to eliminate ambiguity during data insertion and validation.

The project takes aim to reveal the significant information security criteria in a smart grid that would impact consumer's trust and satisfaction towards the entire system.

ACKNOWLEDGMENT

The authors wish to thank Program GCOE, Graduate School of Science and Technology Meiji University and Japanese Government (MONBUKAGAKUSHO: MEXT) Scholarship for the sponsor, Professor Mimura Masayasu and Professor Sugihara Kokichi from GCOE Program of Meiji University for the mentor. The authors also express gratitude to Dr. Zainol Mustafa and Dr. Rika Fatimah from School of Mathematical Sciences National University of Malaysia for discussion and ideas.